\documentclass{optica-article}

\journal{opticajournal} % for journals or Optica Open

\articletype{Research Article}
\usepackage[export]{adjustbox}
\usepackage{lineno}
\usepackage{graphicx}
\usepackage{subcaption}
\usepackage{amsmath}
\begin{document}

\title{Development of a pyramidal magneto-optical trap for pressure sensing application}

\author{S. Supakar,\authormark{1,*}
 Vivek Singh,\authormark{1,3} Y. Pavan Kumar,\authormark{2} \\S. K. Tiwari,\authormark{2} C. Mukherjee,\authormark{2,3} M. P. Kamath,\authormark{2} \\V. B. Tiwari,\authormark{1,3} and S. R. Mishra\authormark{1,3}}

\address{\authormark{1}Laser Physics Applications Division, Raja Ramanna Centre for Advanced Technology, Indore 452013, India\\
 \authormark{2}High Energy Lasers and Optics Section, Raja Ramanna Centre for Advanced Technology, Indore 452013, India\\
\authormark{3}Homi Bhabha National Institute, Training School Complex, Anushakti Nagar, Mumbai 400094, India}

\email{\authormark{*}subhajits@rrcat.gov.in} %% email address is required; see note below about the corresponding author designation

% use {asbstract*} to suppress the copyright line. Copyright information will be added in production

\begin{abstract*} 
Here, we report the development and working of a compact rubidium (Rb) atom magneto-optical trap (MOT) operated with a hollow pyramidal mirror and a single laser beam. This type of compact MOT is suitable for developing portable atom-optic devices, as it works with less number of optical components as compared to conventional MOT setup. The application of this compact MOT setup for pressure sensing has been demonstrated.
\end{abstract*}

\section{Introduction}
A magneto-optical trap (MOT) is a work-horse setup used for generating samples of cold atoms for their applications in basic science experiments and developing various atom based quantum technologies. Using cold atoms from a MOT, several atom-optic precision measurement devices have been developed, which have shown measurement accuracy comparable or better than existing other devices. These atom-optic devices include atomic clocks \cite{Ludlow}, cold atom gravimeters \cite{Bidel,Hinton,Bodart}, cold atom gyroscopes\cite{Riedl}, cold atoms UHV sensors \cite{Supakar},  etc. Besides that, MOT also serves as a source for realization of qubits, each made of single atom, for quantum information processing application \cite{Frese}. World-wide, different research groups have made considerable efforts to miniaturize the MOT setup\cite{rushton, Earl}, by minimizing the number of laser sources as well minimizing the optical and mechanical components (e.g. beam splitters, retardation plates, mirrors, etc). For example, the need of two sets of frequency-stabilized lasers (called cooling and re-pumping lasers) required at different wavelengths to operate a $^{87}$Rb-MOT, can be fulfilled by using a single laser with current modulation to simultaneously generate the cooling and re-pumping emissions from the same laser source \cite{Myatt, Wiegand}. Alternatively, a single laser with an electro-optic modulator (EOM) in the output can generate the cooling and re-pumping radiation in the same beam \cite{Suptitz, Harada}.  In order to minimize the number of optical and mechanical components required for MOT operation, use of some specially designed optical elements in suitable configuration have also been proposed. These include the use of a mirror surface in refection geometry \cite{Wildermuth,vivek}, use of a grating \cite{Arnold, Nshii, Eckel, Eckel1, Eckel2, Seo, Lee1, Imhof, McGilligan, Lee2, Cotter, Duan}, and use of a pyramidal mirror \cite{Foot, Wu, Kohel, Williamson, Pollock, Pollock1, Lee} in the setup for MOT. The latter two cases use a single cooling laser beam, which  provides operational ease and compactness to the setup. 

Here we report the development and working of a compact and simple MOT setup for $^{87}$Rb atoms using a pyramidal mirror. We use a single ECDL laser system with an EOM in the output to obtain both cooling and re-pumping laser emissions in a single beam. This beam (called MOT beam) when enters into a home made hollow pyramidal mirror, the required six MOT beams are generated inside the hollow region of pyramidal mirror. The fabrication of the hollow  pyramidal mirror has been discussed and operation of the magneto-optical trap (MOT) using this mirror has been demonstrated. Nearly 7.9 $\times$ 10$^6$ $^{87}$Rb atoms have been trapped in the pyramidal-MOT. This development is important for making transportable cold atoms source. The initial use of this MOT setup for sensing ultrahigh vacuum (UHV) in the chamber has been demonstrated.

% \begin{figure}[h]
% \centering
% \includegraphics[scale=0.37]{angle_measurement.png}
% \caption{Schematic of the experimental setup of the measurement of the angles between the adjacent faces of the pyramidal mirror. (a) one of the reflecting face A of the pyramidal mirror is normal to the incident beam. (b) the PM is translated in horizontal direction such that laser beam is made incident on its face B (the face adjacent to the face A). (c) photograph of the experimental setup. L: He-Ne laser, AP: aperture, BS: cube beam splitter, BD: beam dump, ND: neutral density filters, FL: focusing lens with f=100mm, CCD: camera, ROS: reference optical system.}
% \label{angle-measurement}
% \end{figure}

\section{Fabrication of hollow pyramidal mirror }

\begin{figure}[h]
\centering
\includegraphics[scale=0.95]{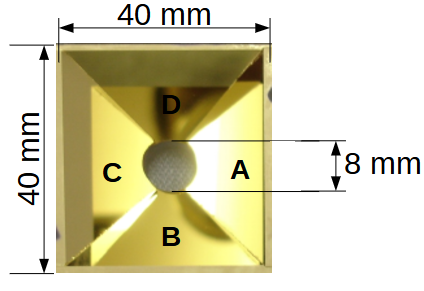}
\caption{Photograph of the gold-coated hollow pyramidal mirror for use in compact single beam operated magneto-optical trap (MOT) setup.}
\label{mirror}
\end{figure}

 The photograph of hollow pyramidal mirror (PM) is shown in Fig.\ref{mirror}. It has the base size of 40 mm × 40 mm and height of 30 mm. There is a hole of ~ 8 mm diameter at the apex of this pyramidal mirror. It is custom-designed and fabricated in-house for our magneto-optical trap (MOT) setup. It is made of four identical triangular faces, in which adjacent faces made an angle of $120^{\circ}$ between them and the opposite faces made an angle of $90^{\circ}$ with each other at vertex point. These angles have been measured using laser interferometry techniques \cite{Malacara} and using auto-collimators \cite{Yoder}. Among these four faces, two opposite faces A and C are made using two right angle glass prisms, and the other two faces B and D are made using two corner cube glass prisms, Fig.\ref{mirror}. All the four faces facing inside the pyramidal mirror were polished using the chemical-mechanical polishing process before joining them. An auto-collimator was used to align faces of the prisms before they were cemented. The accuracy of angles on joining the prisms faces was maintained at the level of 10 arc sec. After joining all the faces, a hole of 8 mm diameter at vertex of this hollow pyramid was made by applying the optical trepanning technique. In optical trepanning, a hollow cylindrical tool with required external diameter is rotated on the optical component with a downward pressure and loose abrasive (Aluminum oxide power) is fed for cutting. We have used 50 mm long trepanning tool made of brass with external diameter of 7.9 mm and wall thickness of 1 mm. The purpose of the making this hole was to allow the passage for trapped cold atoms out side the pyramidal mirror to perform further experiments. One of these experiments could be to allow fall of atoms in earth's gravitational field perform Raman pulse atom interferometry for accurate measurement of earth's gravitational acceleration (g). \\
 
Subsequent to joining of faces and cutting of hole at the apex, the inner polished surfaces of pyramidal mirror were coated for achieving reflectivity from the inner faces. For this, a 15 nm thick chromium (Cr) layer was deposited on all the four inner faces of the pyramidal element before coating these surfaces with gold. This is necessary to improve the adhesion of gold layer on glass surface of the pyramid. Then a 200 nm thick gold (Au) layer was deposited on these chromium coated surfaces for achieving the desired high optical reflectivity of 780 nm laser beam from these surfaces. A pulsed DC magnetron sputtering system was  used to deposit both the layers, chromium and gold, on the inner faces of the hollow pyramid. The finally finished pyramidal mirror (PM) was installed on a suitable mount of stainless steel (SS) before it was inserted in an ultra-high vacuum (UHV) chamber for MOT operation. The background pressure of $\approx\,1\,\times\,10^{-8}\,$ Torr was generated in this chamber using a turbo molecular pump (TMP 70 L/s) and a sputter ion pump (SIP 20 L/s). 

\section{Pyramidal MOT setup}
\begin{figure}[h]
\centering
\includegraphics[scale=0.4]{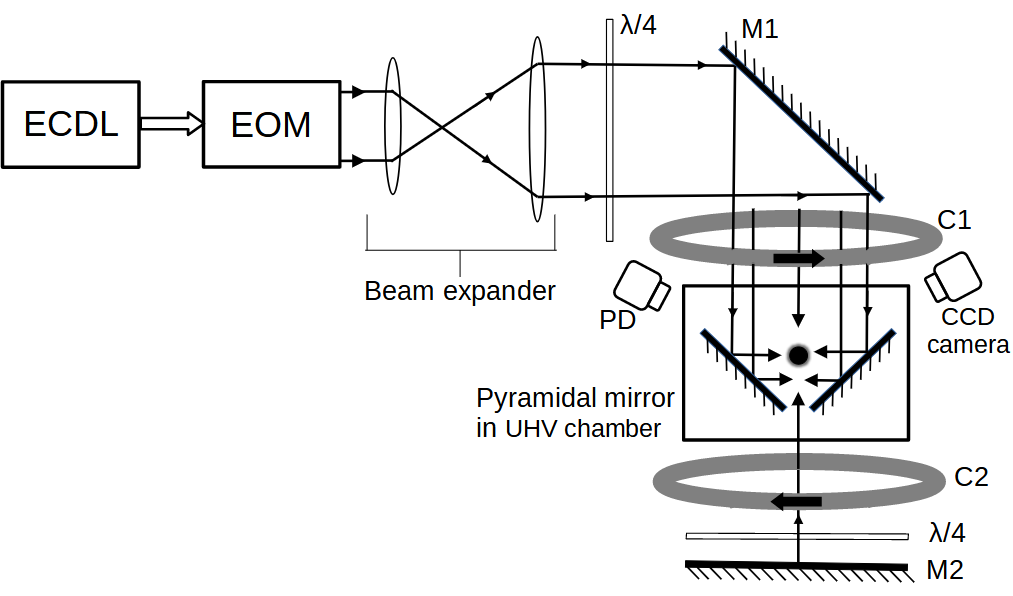}
\caption{Schematic of pyramidal MOT setup. ECDL: External cavity diode laser, EOM: Electro-optic modulator, C1 and C2: quadrupole coils, M1 and M2: Mirrors, PD: Photodiode.}
\label{setup}
\end{figure}

Fig. \ref{setup} shows an schematic of the experimental setup for magneto-optical trap (MOT) using the in-house fabricated gold coated pyramidal mirror. We need only one laser beam incident on pyramidal mirror for operation of MOT for $^{87}$Rb atoms. From this incident single laser beam,  the pyramidal mirror generates the required MOT beams for three-dimensional trapping of atoms after reflection of the incident beam from the its surfaces and from a mirror M2, as shown in Fig.2. This reduces the requirement of several optical elements, such as beam splitters and mirrors, for generating the six MOT beams required for a MOT setup.

To generate both cooling and re-pumping radiations in the same single MOT beam, we passed the output laser beam from an ECDL through an electro-optic modulator (EOM) device. For this the ECDL output beam was locked to the cooling transition of $^{87}$Rb atom before passing through the EOM. We have used an ECDL (Make: TOPTICA, DLC-Pro) operating at 780 nm wavelength. The maximum power of the ECDL was $\sim$ 100 mW. We used an EOM (Make: Qubig, Germany, Model: EO-Rb87-6.6G) which was operated using rf field at frequency of $\nu_{RF}$ = 6.56 GHz. When ECDL output was passed through the EOM, the resulting output beam had both cooling and re-pumping emissions. This output beam served as MOT beam for our experiments. In the output beam from EOM, the cooling beam at frequency $\nu_L$ and re-pumping laser beam at frequency $\nu_L + \nu_{RF}$ were obtained in co-propagating geometry. In the EOM output beam, the distribution of power in cooling and re-pumping parts was distributed with $\sim60\%$ and $\sim20\%$ of the total output beam power. The remaining 20 \% power at frequency  $\nu_L - \nu_{RF}$ in the EOM output beam was un-utilized. In order to achieve these values, the EOM device was operated at power of $\sim32$ dBm.  \\

\begin{figure}
\centering
    \begin{subfigure}[b]{0.5\textwidth}            
            \includegraphics[width=\textwidth]{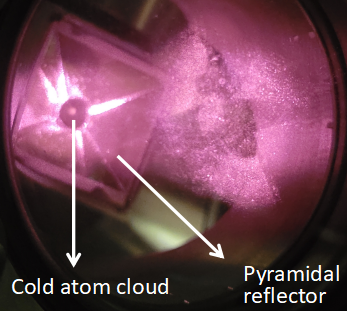}
            \caption{}
            \label{cloud_img}
    \end{subfigure}%
     %add desired spacing between images, e. g. ~, \quad, \qquad etc.
     ~
    \begin{subfigure}[b]{0.6\textwidth}
            \centering
            \includegraphics[width=\textwidth]{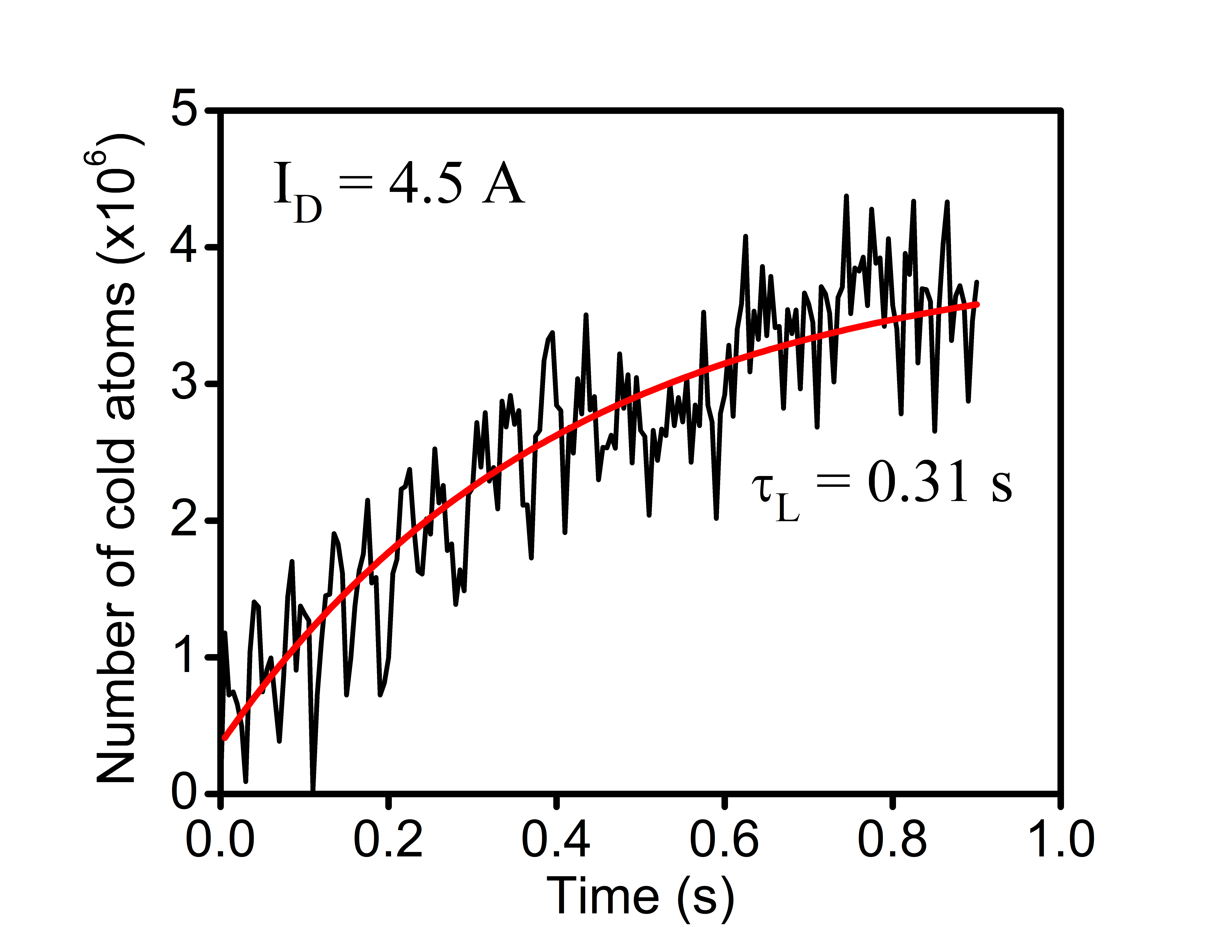}
            \caption{}
            \label{loading}
    \end{subfigure}
    \caption{(a) Photograph of cold atom cloud of $^{87}$Rb atoms trapped in the pyramidal MOT. (b) The observed MOT loading curve which shows the temporal variation in fluorescence signal (proportional to number of atoms) from atoms in MOT.} 
\end{figure}

 The output beam from EOM was expanded using a beam expander and then this linearly polarized beam was passed through a quarter-wave plate for converting its polarization to circular polarization. The diameter of beam was ~ 20 mm after the expansion. Before entry into the UHV chamber, the power of the MOT beam was 12 mW, with 7.2 mW in cooling part and 2.4 mW in repuming part. With the use of a retro-reflecting plane mirror (M2) in the back side of the pyramidal mirror, the six MOT beams configuration was achieved with single input beam, as shown in Fig.\ref{setup}. Each reflection reverses the helicity of the circularly polarized beam so that the desired $\sigma^-\,-\,\sigma^+$ polarization configuration was produced along the three orthogonal axes. 
 
 The Rb-source (a combination of Rb-getter dispensers ) was installed in the UHV chamber using a two-pin vacuum feed-throgh. A variable dc current, called dispenser current (I$_D$), was supplied to this Rb-source for generation of Rb vapor in the chamber. The dispenser source was positioned at a distance of $\sim$ 10 cm from the MOT position. A pair of DC current carrying coils (in anti-Helmholtz like configuration) was installed out side the UHV chamber for generating the required quadrupole magnetic field for the MOT. These coils were placed such that the coils axis was parallel to the input laser beam direction. The coils were supplied a DC current of 5.3 A to produce axial magnetic field gradient of $\sim$ 11 G/cm. Another pair of DC bias coils in Helmholtz like configuration was added in order to manipulate the zero position of the quadrupole magnetic field. This was necessary to ascertain the maximum number of atoms trapped in the MOT. 
 
 A photograph of the cloud of $^{87}$Rb atoms trapped in pyramidal MOT is shown in Fig. \ref{cloud_img}. Nearly $3.5\times10^6$ atoms have been trapped in the MOT at the dispenser current of I$_D =$ 4.5 A. To study the loading behavior of the MOT, fluorescence from the atoms in the MOT was collected on a photodiode (PD). The photodiode signal was calibrated for the number of atoms using a CCD camera. The CCD camera was used to capture fluorescence image of the atom cloud and estimate the number of atoms in the MOT \cite{Mishra}. Fig. \ref{loading} shows the MOT loading curve at dispenser current of 4.5 A and cooling beam detuning of -15 MHz (i.e. red side of cooling transition). The red curve in Fig. \ref{loading} is the best fit of the loading equation $N(t) = N_s\left[1-exp(-t/\tau_{L})\right]$ to the experimentally observed MOT loading curve, where N(t) is number of atoms at time t and N$_S$ is the saturated number in the MOT. From the best fit to the experimental data in Fig. \ref{loading}, we get N$_S$ = $3.5\times10^6$ and $\tau_L = 0.31\,$s.

\section{Pyramidal MOT : Application to pressure sensing}

Here, we discuss our measurements of pressure using pyramidal MOT loading characteristics. The MOT loading curve in Fig. \ref{loading} shows the observed loading behaviour of our MOT. Here we measured the variation in number of atoms in the MOT as function of time. As can be seen in the Fig. \ref{loading}, the initial rise in number of atoms is exponential before reaching to the saturated number of atoms. 

The loading of a MOT, neglecting the intra-trap collisional loss rate, is governed by a rate equation,
\begin{equation}
    \frac{dN}{dt} = R-\frac{N}{\tau_L}
\end{equation}
where N is the number of atoms in MOT at any time t and R is the MOT loading rate from the Rb-vapour in the background. The solution of equation (1) gives the MOT loading equation as,  
\begin{equation}
    N(t) = N_s\left[1-exp(-t/\tau_{L})\right],
\end{equation}
where
\begin{equation}
    N_S = R\tau_{L} = \alpha P_{Rb}\tau_{L}.
\end{equation}

Here, N$_S$ represents the final saturated number of atoms in the MOT and R=$\alpha$$P_{Rb}$ represents the MOT loading rate. The time constant $\tau_L$ is called MOT loading time. The inverse of $\tau_L$, i.e. $\gamma$ = $\frac{1}{\tau_L}$, represents the loss rate of atoms in the MOT. Neglecting intra-trap collisions, the loss of atoms from the MOT is mainly due to collisions of MOT atoms with untrapped atoms and molecules present in the surroundings of MOT cloud. The partial pressure of a gas in the chamber determines the collisional loss rate of MOT atoms due to that particular gas. Therefore in a vapor loaded MOT, the MOT loss rate can be expressed as sum of loss rate due to free Rb vapor atoms and loss rate due to other gases in the chamber. Thus, we can write the total loss rate as, 
\begin{equation}
    \gamma=\frac{1}{\tau_{L}} = {\gamma_{Rb} + \gamma_{bg}}= \beta P_{Rb}+\gamma_{bg},
\end{equation} 
where the term $\beta P_{Rb}$ is the loss rate of MOT atom due to their collisions with hot Rb atoms in the vapor, with $ P_{Rb}$ being the partial pressure of Rb atomic vapor in the chamber. The term $\gamma_{bg}$ represents the MOT loss rate due to collision of other atoms/molecules (excluding Rb atoms) in the chamber with Rb atoms in MOT.

 After using equation (3) and (4), the relation between N$_S$ and $\tau_L$ is given by
\begin{equation}
    N_S = \frac{\alpha}{\beta}(1-\gamma_{bg} \tau_L).
\end{equation}
Thus, by measuring the variation of N$_S$ and $\tau_L$ for different Rb pressure values (by varying the dispenser current), it is possible to estimate the Rb pressure (from $\alpha/\beta$) as well as non-Rb gases pressure (from $\gamma_{bg}$) in the chamber. 

\begin{table}[h!]
\centering
\caption{ The measured values of $\tau_{L}$, N$_S$ and P$_{Rb}$ for different values of Rb-dispenser current ( keeping other parameter unchanged during experiments).}
\begin{tabular}{|c|c|c|c|c|}
\hline
S. No. & Disp. current I$_D$ (A) & $\tau_{L}$ (s) & N$_S$ & P$_{Rb}$ (Torr)\\
\hline
1 & 4.5  & 0.31 & $3.5\,\times$ 10$^6$ & $1.2\,\times10^{-8}$\\
\hline
2 & 5.0  & 0.28 & $4.0\,\times$ 10$^6$ & $1.6\,\times10^{-8}$\\
\hline
3 & 5.5  & 0.23 & $7.9\,\times$ 10$^6$ & $3.7\,\times10^{-8}$\\
\hline
\end{tabular}
\label{Tab:1}

\end{table}
In our experiment, the value of  $\alpha/\beta$ and $\gamma_{bg}$ are estimated by fitting the data points listed in Table 1. From the fit, we obtain $\gamma_{bg}$ = 2.6 s$^{-1}$ and $\alpha/\beta$ = 2.1$\times$ 10$^7$. This value of $\gamma_{bg}$ can be used to estimate the background pressure (P) due to non-Rb gases in the chamber, using the relation $\gamma_{bg}$/P = 4.9 $\times$10$^7$ Torr$^{-1}$ s$^{-1}$ \cite{Arpo,Moore}. Similarly, the value of $\alpha/\beta$ can be used to estimate the value of $\alpha$, using the value of $\beta$ = 4.4 $\times$10$^7$ Torr$^{-1}$ s$^{-1}$ available in the literature \cite{Arpo,Moore}. After knowing the value of $\alpha$, the partial pressure due to Rb vapor ($ P_{Rb}$) can be estimated by using the relation $N_S =\alpha P_{Rb}\tau_{L}$. Here we obtain the value of $\alpha$ as $\alpha$ = 9.2 $\times$ 10 $^{14}$ Torr$^{-1}$ s$^{-1}$ which gives partial pressure due to Rb vapor as 1.2 $\times$ 10$^{-8}$ Torr at the dispenser current of 4.5 A. The value of partial pressure of Rb vapor estimated by this this method at different values of dispenser current is given in Table 1. Using the value of $\gamma_{bg}$, we estimate the background (non-Rb) pressure (P) in our chamber as $(5.3 \pm0.9)\times$ 10$^{-8}$ Torr, using the relation $\gamma_{bg}$/P = (1/k) = 4.9 $\times$10$^7$ Torr$^{-1}$ s$^{-1}$, assuming only hydrogen ($H_2$) gas in the background \cite{Arpo,Moore}. This estimated non-Rb background gases pressure value is slightly higher than the value of pressure ($\approx\,1\,\times\,10^{-8}\,$ Torr) read by the ion gauge in the setup. This difference could be due to the conductance of the pipe used for attaching ion gauge to the chamber. Further, the estimation of the background gases pressure from the $\gamma_{bg}$ is dependent on MOT parameters and the value of k. The value of k is also dependent on the composition of the background gases, and it can be higher by a factor of 2 (or more) if other gases (He, Ar and $H_2$O) are present in significant fraction in the MOT chamber \cite{Supakar}. Further work is in progress to ascertain the background non-Rb pressure in the chamber more accurately. 

\section{Conclusion}
We have demonstrated the working of a magneto-optical trap with a pyramidal mirror for pressure sensing application. The pyramidal mirror was fabricated in-house using four gold-coated prism surfaces. Using this MOT setup, the background Rb and non-Rb gases pressure values were estimated from the MOT loading data. This compact MOT setup is a useful development for making field deployable cold atom based UHV pressure sensor.

%%Appendix

%%Use section* for acknowledgements
\section*{Acknowledgement}
We acknowledge the help extended by S. P. Ram, S. Singh, K. Bhardwaj, D. Mewara and S. Bhardwaj during the experiments. We are also thankful to N. S. Benerji and K. S. Bindra for the technical support in this work. Authors are also grateful to S. V. Nakhe for fruitful discussions and suggestions.

%%use \balance somewhere in the left column of the last page to balance the two columns in the end page

%%References section
%\begin{thebibliography}{99} 
%\bibitem{latexcompanion} 
%Michel Goossens, Frank Mittelbach, and Alexander Samarin. 
%\newblock {\em The \LaTeX\ Companion}. 
%Addison-Wesley, Reading, Massachusetts, 1993.
%\end{thebibliography}
\medskip
\noindent The authors declare no conflict among themselves.

%%%%%%%%%%%%%%%%%%%%%%%%%%  body  %%%%%%%%%%%%%%%%%%%%%%%%%%

\bibliography{reference}

\begin{thebibliography}{10}
\newcommand{\enquote}[1]{``#1''}

\bibitem{Ludlow}
A.~D. Ludlow, M.~M. Boyd, J.~Ye, E.~Peik, and P.~O. Schmidt, \enquote{Optical atomic clocks,} {\protect\JournalTitle{Rev. Mod. Phys.}} \textbf{87}, 637--701 (2015).

\bibitem{Bidel}
Y.~Bidel, O.~Carraz, R.~Charrière, M.~Cadoret, N.~Zahzam, and A.~Bresson, \enquote{Compact cold atom gravimeter for field applications,} {\protect\JournalTitle{Applied Physics Letters}} \textbf{102} (2013).

\bibitem{Hinton}
A.~Hinton, M.~Perea-Ortiz, J.~Winch, J.~Briggs, S.~Freer, D.~Moustoukas, S.~Powell-Gill, C.~Squire, A.~Lamb, C.~Rammeloo, B.~Stray, G.~Voulazeris, L.~Zhu, A.~Kaushik, Y.-H. Lien, A.~Niggebaum, A.~Rodgers, A.~Stabrawa, D.~Boddice, S.~R. Plant, G.~W. Tuckwell, K.~Bongs, N.~Metje, and M.~Holynski, \enquote{A portable magneto-optical trap with prospects for atom interferometry in civil engineering,} {\protect\JournalTitle{Philosophical Transactions of the Royal Society A: Mathematical, Physical and Engineering Sciences}} \textbf{375}, 20160238 (2017).

\bibitem{Bodart}
Q.~Bodart, S.~Merlet, N.~Malossi, F.~P. Dos~Santos, P.~Bouyer, and A.~Landragin, \enquote{{A cold atom pyramidal gravimeter with a single laser beam},} {\protect\JournalTitle{Applied Physics Letters}} \textbf{96}, 134101 (2010).

\bibitem{Riedl}
S.~Riedl, G.~W. Hoth, B.~Pelle, J.~Kitching, and E.~A. Donley, \enquote{Compact atom-interferometer gyroscope based on an expanding ball of atoms,} {\protect\JournalTitle{Journal of Physics: Conference Series}} \textbf{723}, 012058 (2016).

\bibitem{Supakar}
S.~Supakar, V.~Singh, V.~B. Tiwari, and S.~R. Mishra, \enquote{{Ultrahigh vacuum pressure measurement using magneto-optical trap on atom chip},} {\protect\JournalTitle{Journal of Applied Physics}} \textbf{134}, 024403 (2023).

\bibitem{Frese}
D.~Frese, B.~Ueberholz, S.~Kuhr, W.~Alt, D.~Schrader, V.~Gomer, and D.~Meschede, \enquote{Single atoms in an optical dipole trap: Towards a deterministic source of cold atoms,} {\protect\JournalTitle{Phys. Rev. Lett.}} \textbf{85}, 3777--3780 (2000).

\bibitem{rushton}
J.~A. Rushton, M.~Aldous, and M.~D. Himsworth, \enquote{Contributed review: The feasibility of a fully miniaturized magneto-optical trap for portable ultracold quantum technology,} {\protect\JournalTitle{Rev. of Sci. Instruments}} \textbf{85}, 121501 (2014).

\bibitem{Earl}
L.~Earl, J.~Vovrosh, M.~Wright, D.~Roberts, J.~Winch, M.~Perea-Ortiz, A.~Lamb, F.~Hayati, P.~Griffin, N.~Metje, K.~Bongs, and M.~Holynski, \enquote{Demonstration of a compact magneto-optical trap on an unstaffed aerial vehicle,} {\protect\JournalTitle{Atoms}} \textbf{10}, 32 (2022).

\bibitem{Myatt}
C.~J. Myatt, N.~R. Newbury, and C.~E. Wieman, \enquote{Simplified atom trap by using direct microwave modulation of a diode laser,} {\protect\JournalTitle{Opt. Lett.}} \textbf{18}, 649--651 (1993).

\bibitem{Wiegand}
B.~Wiegand, B.~Leykauf, K.~Döringshoff, Y.~D. Gupta, A.~Peters, and M.~Krutzik, \enquote{{A single-laser alternating-frequency magneto-optical trap},} {\protect\JournalTitle{Review of Scientific Instruments}} \textbf{90} (2019). 103202.

\bibitem{Suptitz}
W.~S\"{u}ptitz, G.~Wokurka, F.~Strauch, P.~Kohns, and W.~Ertmer, \enquote{Simultaneous cooling and trapping of 85rb and 87rb in a magneto-optical trap,} {\protect\JournalTitle{Opt. Lett.}} \textbf{19}, 1571--1573 (1994).

\bibitem{Harada}
K.~Harada, T.~Aoki, S.~Ezure, K.~Kato, T.~Hayamizu, H.~Kawamura, T.~Inoue, H.~Arikawa, T.~Ishikawa, T.~Aoki, A.~Uchiyama, K.~Sakamoto, S.~Ito, M.~Itoh, S.~Ando, A.~Hatakeyama, K.~Hatanaka, K.~Imai, T.~Murakami, H.~S. Nataraj, Y.~Shimizu, T.~Sato, T.~Wakasa, H.~P. Yoshida, and Y.~Sakemi, \enquote{Laser frequency locking with 46\&\#x2009;\&\#x2009;ghz offset using an electro-optic modulator for magneto-optical trapping of francium atoms,} {\protect\JournalTitle{Appl. Opt.}} \textbf{55}, 1164--1169 (2016).

\bibitem{Wildermuth}
S.~Wildermuth, P.~Kr\"uger, C.~Becker, M.~Brajdic, S.~Haupt, A.~Kasper, R.~Folman, and J.~Schmiedmayer, \enquote{Optimized magneto-optical trap for experiments with ultracold atoms near surfaces,} {\protect\JournalTitle{Phys. Rev. A}} \textbf{69}, 030901 (2004).

\bibitem{vivek}
V.~Singh, V.~B. Tiwari, K.~A.~P. Singh, and S.~R. Mishra, \enquote{On loading of a magneto-optical trap on an atom-chip with u-wire quadrupole field,} {\protect\JournalTitle{Journal of Modern Optics}} \textbf{65}, 2332--2338 (2018).

\bibitem{Arnold}
M.~Vangeleyn, P.~F. Griffin, E.~Riis, and A.~S. Arnold, \enquote{Laser cooling with a single laser beam and a planar diffractor,} {\protect\JournalTitle{Opt. Lett.}} \textbf{35}, 3453--3455 (2010).

\bibitem{Nshii}
C.~C. Nshii, M.~Vangeleyn, J.~P. Cotter, P.~F. Griffin, E.~A. Hinds, C.~N. Ironside, P.~See, A.~G. Sinclair, E.~Riis, and A.~S. Arnold, \enquote{A surface-patterned chip as a strong source of ultracold atoms for quantum technologies,} {\protect\JournalTitle{Nature Nanotechnology}} \textbf{8}, 321--324 (2013).

\bibitem{Eckel}
A.~Sitaram, P.~K. Elgee, G.~K. Campbell, N.~N. Klimov, S.~Eckel, and D.~S. Barker, \enquote{Confinement of an alkaline-earth element in a grating magneto-optical trap,} {\protect\JournalTitle{Rev. of Sci. Instrum.}} \textbf{91}, 103202 (2020).

\bibitem{Eckel1}
D.~Barker, E.~Norrgard, N.~Klimov, J.~Fedchak, J.~Scherschligt, and S.~Eckel, \enquote{Single-beam zeeman slower and magneto-optical trap using a nanofabricated grating,} {\protect\JournalTitle{Phys. Rev. Appl.}} \textbf{11}, 064023 (2019).

\bibitem{Eckel2}
W.~R. McGehee, W.~Zhu, D.~S. Barker, D.~Westly, A.~Yulaev, N.~Klimov, A.~Agrawal, S.~Eckel, V.~Aksyuk, and J.~J. McClelland, \enquote{Magneto-optical trapping using planar optics,} {\protect\JournalTitle{New Journal of Physics}} \textbf{23}, 013021 (2021).

\bibitem{Seo}
S.~Seo, J.~H. Lee, S.-B. Lee, S.~E. Park, M.~H. Seo, J.~Park, T.~Y. Kwon, and H.-G. Hong, \enquote{Maximized atom number for a grating magneto-optical trap via machine-learning assisted parameter optimization,} {\protect\JournalTitle{Opt. Express}} \textbf{29}, 35623--35639 (2021).

\bibitem{Lee1}
J.~Lee, J.~A. Grover, L.~A. Orozco, and S.~L. Rolston, \enquote{Sub-doppler cooling of neutral atoms in a grating magneto-optical trap,} {\protect\JournalTitle{J. Opt. Soc. Am. B}} \textbf{30}, 2869--2874 (2013).

\bibitem{Imhof}
E.~Imhof, B.~K. Stuhl, B.~Kasch, B.~Kroese, S.~E. Olson, and M.~B. Squires, \enquote{Two-dimensional grating magneto-optical trap,} {\protect\JournalTitle{Phys. Rev. A}} \textbf{96}, 033636 (2017).

\bibitem{McGilligan}
J.~P. McGilligan, P.~F. Griffin, R.~Elvin, S.~J. Ingleby, E.~Riis, and A.~S. Arnold, \enquote{Grating chips for quantum technologies,} {\protect\JournalTitle{Scientific Reports}} \textbf{7}, 384 (2017).

\bibitem{Lee2}
J.~Lee, R.~Ding, J.~Christensen, R.~R. Rosenthal, A.~Ison, D.~P. Gillund, D.~Bossert, K.~H. Fuerschbach, W.~Kindel, P.~S. Finnegan, J.~R. Wendt, M.~Gehl, A.~Kodigala, H.~McGuinness, C.~A. Walker, S.~A. Kemme, A.~Lentine, G.~Biedermann, and P.~D.~D. Schwindt, \enquote{A compact cold-atom interferometer with a high data-rate grating magneto-optical trap and a photonic-integrated-circuit-compatible laser system,} {\protect\JournalTitle{Nature Communications}} \textbf{13}, 5131 (2022).

\bibitem{Cotter}
J.~P. Cotter, J.~P. McGilligan, P.~F. Griffin, I.~M. Rabey, K.~Docherty, E.~Riis, A.~S. Arnold, and E.~A. Hinds, \enquote{Design and fabrication of diffractive atom chips for laser cooling and trapping,} {\protect\JournalTitle{App. Phys. B}} \textbf{122}, 172 (2016).

\bibitem{Duan}
J.~Duan, X.~Liu, Y.~Zhou, X.-B. Xu, L.~Chen, C.-L. Zou, Z.~Zhu, Z.~Yu, N.~Ru, and J.~Qu, \enquote{High diffraction efficiency grating atom chip for magneto-optical trap,} {\protect\JournalTitle{Optics Communications}} \textbf{513}, 128087 (2022).

\bibitem{Foot}
J.~Arlt, O.~Maragò, S.~Webster, S.~Hopkins, and C.~Foot, \enquote{A pyramidal magneto-optical trap as a source of slow atoms,} {\protect\JournalTitle{Optics Communications}} \textbf{157}, 303--309 (1998).

\bibitem{Wu}
X.~Wu, F.~Zi, J.~Dudley, R.~J. Bilotta, P.~Canoza, and H.~M\"{u}ller, \enquote{Multiaxis atom interferometry with a single-diode laser and a pyramidal magneto-optical trap,} {\protect\JournalTitle{Optica}} \textbf{4}, 1545--1551 (2017).

\bibitem{Kohel}
J.~M. Kohel, J.~Ramirez-Serrano, R.~J. Thompson, L.~Maleki, J.~L. Bliss, and K.~G. Libbrecht, \enquote{Generation of an intense cold-atom beam from a pyramidal magneto-optical trap: experiment and simulation,} {\protect\JournalTitle{J. Opt. Soc. Am. B}} \textbf{20}, 1161--1168 (2003).

\bibitem{Williamson}
R.~S.~W. III, P.~A. Voytas, R.~T. Newell, and T.~Walker, \enquote{A magneto-optical trap loaded from a pyramidal funnel,} {\protect\JournalTitle{Opt. Express}} \textbf{3}, 111--117 (1998).

\bibitem{Pollock}
S.~Pollock, J.~P. Cotter, A.~Laliotis, F.~Ramirez-Martinez, and E.~A. Hinds, \enquote{Characteristics of integrated magneto-optical traps for atom chips,} {\protect\JournalTitle{New Journal of Physics}} \textbf{13}, 043029 (2011).

\bibitem{Pollock1}
S.~Pollock, J.~P. Cotter, A.~Laliotis, and E.~A. Hinds, \enquote{Integrated magneto-optical traps on a chip using silicon pyramid structures,} {\protect\JournalTitle{Opt. Express}} \textbf{17}, 14109--14114 (2009).

\bibitem{Lee}
K.~I. Lee, J.~A. Kim, H.~R. Noh, and W.~Jhe, \enquote{Single-beam atom trap in a pyramidal and conical hollow mirror,} {\protect\JournalTitle{Opt. Lett.}} \textbf{21}, 1177--1179 (1996).

\bibitem{Malacara}
D.~Malacara and O.~Harris, \enquote{Interferometric measurement of angles,} {\protect\JournalTitle{Appl. Opt.}} \textbf{9}, 1630--1633 (1970).

\bibitem{Yoder}
P.~R. Yoder, E.~R. Schlesinger, and J.~L. Chickvary, \enquote{Active annular-beam laser autocollimator system,} {\protect\JournalTitle{Appl. Opt.}} \textbf{14}, 1890--1895 (1975).

\bibitem{Mishra}
S.~R. Mishra, S.~P. Ram, S.~K. Tiwari, and S.~C. Mehendale, \enquote{Enhanced atom transfer in a double magneto-optical trap setup,} {\protect\JournalTitle{Phys. Rev. A}} \textbf{77}, 065402 (2008).

\bibitem{Arpo}
T.~Arpornthip, C.~A. Sackett, and K.~J. Hughes, \enquote{Vacuum-pressure measurement using a magneto-optical trap,} {\protect\JournalTitle{Phys. Rev. A}} \textbf{85}, 033420 (2012).

\bibitem{Moore}
R.~W.~G. Moore, L.~A. Lee, E.~A. Findlay, L.~Torralbo-Campo, G.~D. Bruce, and D.~Cassettari, \enquote{{Measurement of vacuum pressure with a magneto-optical trap: A pressure-rise method},} {\protect\JournalTitle{Review of Scientific Instruments}} \textbf{86}, 093108 (2015).

\end{thebibliography}

\end{document}